# A Fast Onboard Star-Extraction Algorithm Optimized for the SVOM Visible Telescope

WANG Tianzhou[1,2][†], QIU Yulei[1], CAI Hongbo[1], DENG Jingsong[1]

[1] National Astronomical Observatories, Chinese Academy of Sciences, Beijing 100012, China;
[2] Graduate School of Chinese Academy of Sciences, Beijing 100049, China

The Space multi-band Variable Object Monitor (SVOM) is a proposed Chinese astronomical satellite, dedicated to the detection, localization and measurement of the gamma-ray bursts (GRBs) on the cosmological scale. An efficient algorithm is developed for the purpose of onboard star extraction from the CCD images obtained with the Visible Telescope (VT) on board SVOM. The CCD pixel coordinates of the reference stars will be used to refine the astronomical position of the satellite, which will facilitate trigger rapid ground-based follow-up obsevations of the GRBs. In this algorithm, the image is divided into a number of grid cells and the "global" pixel-value maximum within each cell is taken as the first-guess position of a "bright" star. The correct center postion of a star is then computed using a simple iterative method. Applying two additional strategies, i.e., scanning the image only by even (or odd) lines or in a black-white chess board mode, are proposed to further reduce the time to extract the stars. To examine the efficiency of the above aglorithms, we applied them to the experimental images obtained with a ground-based telescope. We find that the accuracy of the astronomical positioning achieved by our method is comparable to that derived by using the conventional star-extraction method, while the former needs CPU time about 25 times less than the latter. This will significantly improve the performance of the SVOM VT mission.

Received; accepted
doi:
[†]Corresponding author (email: wtz@bao.ac.cn)
Supported by National Basic Research Program of China-973 Program 2009CB824800

China will launch an astronomical satellite-- "Space multi-band Variable Object Monitor" (SVOM)[1,2] -- to detect cosmological gamma-ray bursts (GRBs) and trigger both the onboard and ground-based telescopes for follow-up observations. GRBs appear as transient flashes of gamma-rays in the sky, lasting typically for a couple of seconds and usually being accompanied by prompt and afterglow emissions of X-ray or optical photons. They play important roles in many astrophysical fields[3], including stellar evolution, black hole studies, shock physics and particle acceleration, cosmology, etc..

The SVOM satellite highly emphasizes the coordination between ground-based telescopes and its four scientific instruments onboard, which are the Camera for X- and Gamma-rays (CXG), Gamma-Ray Monitor, Soft X-ray Telescope (SXT), and Visible Telescope (VT). Accurate positioning of GRBs by onboard instruments, e.g. < a few arcsecs, is crucial for ground-based telescopes to measure their redshifts as well as to perform the rapid follow-up observations due to their presence only on relative short timescales.

VT is an optical telescope of 45-cm aperture with two imaging channels, and it is the only onboard instrument of SVOM being able to locate the GRB to be within arcsecond. When a burst is detected and localized (accurate to several arcminutes) by CXG, the platform will immediately slew to its direction towards the follow-up observations with SXT and VT. The first VT image, taken simultaneously with SXT and with short exposure time, will be used to help the GRB localization (accurate to ~0.5 arcminute ) of the latter. Subsequent images will be taken with a longer-time exposure, to detect the optical afterglow of the GRBs and provide a GRB position with an accuray comparable to the sub-arcsecond level.

Owing to a limited capacity of the data transmission of SVOM, the onboard data reduction is usually required. The pixel coordinates of the reference stars extraced in the CCD images will be sent to the ground. Then the extracted reference stars will be used to match with the known catalogues, e.g,, USNO-B, in order to convert the CCD pixel coordinates into the real celestial coordinates.

In this paper, we develop a star-extraction algorithm which can help noticeably reduce on board CPU running time with respect to the conventional method The paper is organized as below: conventional star-extraction methods are briefly reviewed in section 1; our solutions to optimize the star-extracting algorithm are described in section 2; the optimized algorithm is tested on ground-taken experiment images in section 3; the onboard efficiency of the new algorithm is evaluated in section 4; conclusions and discussions are given in section 5.

## 1. An Overview of Conventional Star-Extraction Algorithms

The conventional star-extraction algorithms[4,5] widely used for ground-based data reductions are comprised of two basic steps. The first step is to fit the sky background in order to detect the star pixels from the background. The second is to search for pixels that belong to the same star, and group them together to compute the center coordinates of the star. A typical implementation of such an algorithm is described as below.

1) **Background Estimation**

First, the sky background in each pixel is estimated in order to seperate the object signal from the background signal[4]. The image is divided into many grid cells, each including a number of pixels. In each grid cell, an estimate of the background level is made and is assigned to the center pixel as its nominal value. A relatively "coarse" mesh of the sky background distribution is thus constructed. The value at any other pixel can be obtained from 2-dimensional spline-fitting interpolation of the mesh.

2) **Searching for Star Pixels**

After the interpolated sky background is subtracted from the image, the sky contribution to any star pixel becomes negligible. Only small residues remain in the sky pixels and the mean residue value can be taken as the standard deviation of the background. After convolution with a Gaussian filter, the image is traversed to search for pixels whose values are N times higher than the background standard deviation. Each group of connected high-value pixels are joined together to form an extracted star.

The conventional algorithms described above are time-consuming, although they have the advantage of detecting sources with very low signal-to-noise ratios (SNRs). They are not suitable for satellite onboard data reductions since the onboard computers are usually several hundred times slower than even a normal personal computer available on the ground.

## 2 An Onboard Star-Extraction Algorithm Optimized for the SVOM VT

To transform pixel coordinates in a VT CCD image to celestial ones, we need to extract several reference stars onboard and find their positions. The reference stars to be extracted should be relatively bright in order to have good SNRs. This allows us to develop a fast star-extraction algorithm optimized for the SVOM VT. It consists of two steps as described as follows:

1) **Direct Bright Star Finding**

First the image is divided into ~100-400 grid cells. No evaluation of the sky background will be made. The pixel-value maximum within each cell is found and assumed to indicate a "bright star". We set both an upper threshold and a lower one in order to exclude saturated stars and low-SNR events. The former is chosen as 80% the full-well value of a CCD pixel, while the latter would roughly correspond to a 14 magnitude star for a 15 second exposure of VT. The coordinate of the pixel-value maximum is taken as the first-guess position of the star. The stars can be sorted by the maximum value and only the required number of real bright stars should be retained.

2) **Computing the Correct Star Positions**

The center position of a star is not always coincident with the pixel-value maximum in the star profile. To get the correct position, we do not adopt the time-consuming conventional ways of forming a star profile with all the connected pixels. Instead, we compute the pixel-value centroid of a rectangle region centering at the current position guess and adopt it as the new guess. This is done iteratively till convergence and thus the true center position is found. This procedure usually converges very fast, but if it does not after five steps of iterations we will break the interative loop and the star will be discarded.

The computational complexity of our algorithm is much smaller than that of conventional methods. In the bright-star finding step, the CCD image is traversed only once, so its computational complexity is simply proportional to the total pixel number in an image that is $n$. In our procedure to compute star positions, for each star only hundreds of pixels are visited, negligible compared with $n$. By contrast, the conventional algorithms have a computational complexity proportinal to $n \log(n)$.

We further propose two additional strategies for bright star finding to speed up our star-extraction algorithm. One is to scan the image interlacedly, skipping either odd or even CCD lines and columns (*left* panel of Fig. 1). The other is to divide the image like a black-white chess board, scanning only the black or white patches (*right* panel of Fig. 1). The former will reduce the number of pixels traversed to one fourth of the original algorithm, while the latter will half that number. Since our computational complexity is actually proportional to the number of pixels traversed, by combining both strategies the algorithm should run 8 times faster.

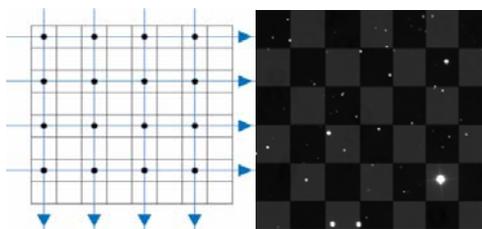

*Fig 1. Two strategies to speed up the star-extraction algorithm, left for interlaced scanning and right for the chess board mode.*

## 3 Testing the Star-Extraction Algorithm

In order to test if the new star-extraction algorithm will work for the future SVOM VT images and how efficient it is, we apply it on astronomical images taken with the 50-cm telescope at the Xinglong Observatory. As shown in Table 1, this telescope is similary to VT in many aspects. In particular, star images produced by both telescopes have nearly the same PSF widths in terms of pixels. Since we are concerned with only bright stars whose SNRs are determined by the source photon counts, the difference in sky brightness between space and ground observations can be ignored. About 200 images of different fields, roughly uniformly distributed in the sky, were taken with the 50-cm telescope.

|              | VT        | 50-cm Telescope |
|--------------|-----------|-----------------|
| Aperture     | 45cm      | 50cm            |
| Focal length | 450cm     | 400cm           |
| FOV          | 21'       | 22'             |
| Resolution   | 0".6      | 1".0            |
| PSF width    | 2.0 pixels| 2.2 pixels      |
| Exposure time| 15s       | 15s             |

*Table1 Specifications of VT and the 50-cm telescope*

As shown by one example of the experimental results (Fig. 2), our algorithm can extract bright stars very efficiently meanwhile avoiding saturated stars. The extracted stars (denoted by *circles*) distribute rather uniformly in the image, suggesting that they are suitable for astronomical positioning.

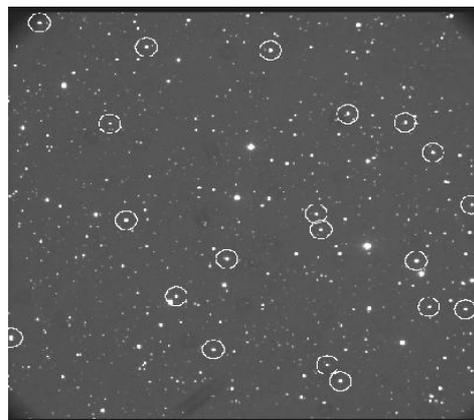

*Fig 2 Stars extracted using the optimized algorithm.*

### 3.1 Bright-Star Extraction Rate vs. Star Brightness

The success of astronomical positioning hinges on the number of bright stars sucessfully extracted. The more stars extracted, the more accurate the matching with a star catalog. Bright stars weigh more than faint ones in the star matching procedure, because they are not susceptible to sky background fluctuations and instrumental noises.

Our tests suggest a sufficiently high star-extraction efficiency of the optimized algorithm. As shown in Fig. 3, for the top 10 brightest stars, 90% can be extracted. The extraction rate gradually decreases as we lower the threshold for bright stars. Nevertheless, as much as 75% of the brightest 60 stars are successfully extracted.

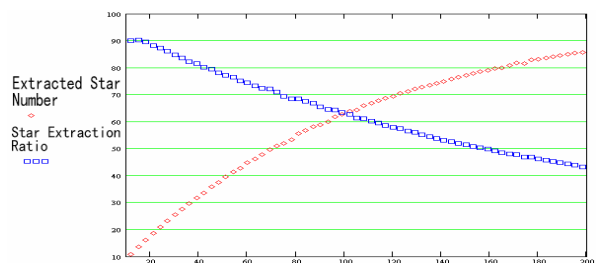

*Fig 3 The bright-star extraction efficiency*

### 3.2 Uniformity of the Positions of Extracted Stars

The position distribution of reference stars in an image is very important for its astronomical positioning. For example, the positioning algorithm may fail if the reference stars are over crowded. If the distribution has a significant bias toward one side of the image, large positioning errors may be introduced.

We evaluate the distribution uniformity of stars extracted with the optimized algorithm, by dividing the CCD image into dozens of equal grid cells and counting the number of extracted stars in each grid cell. As shown in Fig. 4, the extracted stars have a very uniform distribution in the CCD image, making them suitable reference stars for accurate astronomical positioning.

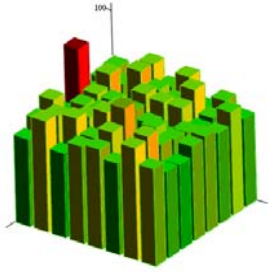
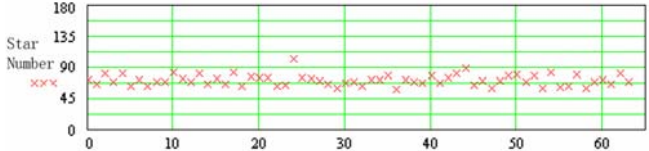

*Fig 4 Evaluation results of bright-star distribution uniformity*

### 3.3 Positioning Accuracy Compared with that of the Conventional Methods

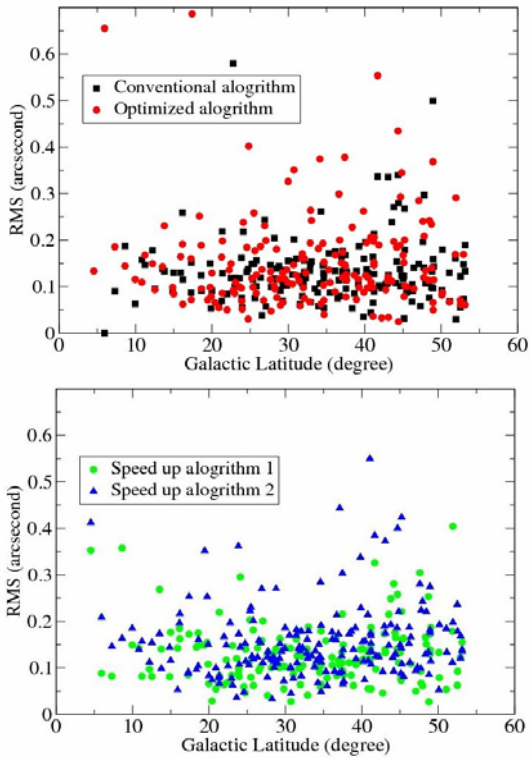

*Fig 5 Positioning accuracy of the optimized method, before (top) and after (bottom) speedup strategies, vs. conventional one.*

A comparison is made in the *top* panel of Fig. 5, of the astronomical positioning accuracy between achieved using the stars extracted with our optimized algorithm and derived with a conventional method. The difference is hardly noticeable. The RMS errors for both methods are below 0.2 arcsecs in most cases, which is sufficient for the GRB localization purpose of VT.

The RMS errors after applying the two additional speedup strategies are plotted in the *bottom* panel of the figure. Both strategies work well, with no significant deterioration in the positioning accuracy.

Although our star-extraction algorithm is much faster than the conventional method, the astronomical positioning achieved by both are almost identical. If neither the interlaced scan nor the chess board mode is used, our error rate of astronimical positioning is close to zero relative to the conventional method. Even when both speedup stratages are applied, the error rate is still as low as about 5%.

We also note that the accuracy of astronomical positioning is not sensitive to the Galactic latitude, despite the fact that star distributions in low Galactic latitudes are much denser than in high Galactic latitudes.

## 4 Evaluation of the Onboard Execution Speed

We first compare the computational complexity between the optimized algorithm and a conventional method, using the Intel VTune[6] performance analysis tool to measure their runtimes. As shown in Table.2, the optimized algorithm only consumes as little as 3.8% the CPU time of the conventional method.

|  | Conventional Algorithm | Optimized Algorithm |
| --- | --- | --- |
| Instruction Retired events | 4,750,153,000 | 367,396,000 |
| Clock ticks events | 13,032,160,000 | 499,104,000 |

*Table2 Runtime performance of a conventional method (left) and the optimized algorithm (right), as analyzed by the Intel VTune.*

To evaluate the expected performance of the optimized algorithm on board SVOM, we implement it on a software simulator of the erc32 processor. Erc32 has been selected as the processor for onboard VT data reductions. Every instruction executed in each part of the processor is monitored. By reference to erc32's hardware manual, the expected onboard runtime of the whole star-extraction procedure is estimated to be 8.4 seconds.

Finally, we test the algorithm on the erc32 experiment board in Shanghai Engineering Center for Microsatellite in order to have a program running environment as similar as possible to the SVOM satellite. The runtime is 7.73 seconds, close to the above estimation.

### 5. Conclusions and discussions

We have developed a fast bright-star extraction algorithm optimized for onboard processing of the VT CCD images of the SVOM satellite. It reduces the CPU time consumed by the star extraction operation for accurate astronomical positioning to about 4% of conventional methods. Its performance regarding the bright-star extraction efficiency, uniformity of the positions of extracted stars, and consequent precision of astronomical positioning is comparable to conventional methods. Testing on a software simulator as well as on the actual hardware equipment verifies that the expected onboard runtime of the algorithm can meet the SVOM engineering requirements.

Our algorithm is simpler than the one adopted by the Ultra-Violet/Optical Telescope on board the Swift satellite[7]. The Swift algorithm identifies all isolated local maxima as stars when traversing the image, while our algorithm only records the global maxima of each grid cells. The Swift one was designed to extract as many stars as possible since the available real-time data transmission rate of Swift is relatively high. But it may falsely claim multiple local maxima for a very bright star which has large extent in pixels. For our specific purpose, i.e. astronomical positioning, only a small number of bright stars are needed. Our algorithm can keep the uniformity of the positions of extracted stars,. This favours accurate astronomical positioning.

*This work has been supported by National Basic Research Program of China – 973 Program 2009CB824800 and by National Natural Science Foundation of China – Grant No. 10878019 and 10673014.*